UEFA EURO 2020: a "pure game of chance"?

Giulia Fedrizzi (1), Luisa Canal (2), Rocco Micciolo (2)

((1) EPSRC Centre for Doctoral Training in Fluid Dynamics, Leeds, UK, (2) Department of Psychology and Cognitive Sciences, University of Trento, Italy)

Category: stat.AP

**Abstract**

We analysed the distribution of the number of goals scored in each of the 51 football matches played in the UEFA EURO 2020 final phase as well as the waiting times between scores (also considering censored times). We found that the Poisson model fits the score data and the exponential distribution fits waiting times quite well. Such a good fit could be considered somewhat counterintuitive and unrealistic given the memoryless property of the exponential model. However, some peculiar features of this study have to be considered: the abilities of the teams were relatively homogeneous; the time span was short; there was no distinction between home and away games; only the total number of goals scored in each game was considered. Although UEFA EURO 2020 can certainly not be considered a "pure game of chance", this competition can be seen as an intriguing example of the pervasive real-world ubiquity of the concept of independence and its ability to encompass complex situations in a single, parsimonious model.

# 1. INTRODUCTION

Statistics in the world of sport has been widely used to calculate a number of descriptive indicators. For example, in a football match one can retrieve information on passing accuracy, possession, free-kicks or corners taken, offsides, and so on. On the other hand, probability is more widely employed in the field of betting since football is an excellent game for different types of gambling.

Being able to model the outcome of a match with a probabilistic model might seem an extremely difficult task, given the complexity of the game and the variety of factors involved (the ability of the teams, the availability of strong and trained players, the tactics defined by the coaches, etc.).

Many authors have focused on models for predicting the outcome of a match (home win, draw, away win) in national leagues taking into account an entire season over a rather long time span (several months or years) (Saraiva et al., 2016; Wheatcroft, 2021). Furthermore, the idea that match statistics might be more informative than goals in terms of making match predictions has become more widespread in recent years (Wheatcroft, 2021).

For example, goals-based team performance covariates were used by Goddard (2005) to forecast win-draw-lose match results. Koopman and Lit (2015) developed a statistical model for the analysis and forecasting of football match results employing time series analysis with intensity coefficients that change stochastically over time. Rue and Salvesen (2000) suggested a Bayesian dynamic generalised linear model to estimate the time-dependent skills of all teams in a league. Wheatcroft (2021) considered the use of observed and predicted match statistics as inputs to forecast the outcomes of football matches.

Other authors have proposed alternative techniques to rank sports teams (Baker and McHale, 2015; Knorr-Held, 2000; Ley et al., 2019).

A wealth of information is available from each football match played. However, if we limit ourselves to the distribution of the number of goals scored, the probabilistic models employed in the literature have been the Poisson (Dixon and Coles, 1997; Karlis and Ntzoufras, 2003; Maher, 1982), negative binomial (Reep et al. 1971; Bittner et al., 2009), extreme-value distributions (Greenhough, 2002; Bittner et al., 2009), or the Weibull count distribution (Boshnakov, 2017). The results are controversial. Some found the Poisson to be adequate, others felt that more parameters needed to be included to improve the fit. Few authors have considered the time between goals.

The recently held final phase of the UEFA EURO 2020 competition gave us the opportunity to evaluate (i) the distribution of the number of goals scored per game (as there was no distinction between home and away games), considering the possibility that some games could last 120 minutes instead of 90, and (ii) the distribution of time between two successive goals taking into account the existence of censored times.

In this paper we want to show how, in the case of UEFA EURO 2020, it is possible to apply a parsimonious probabilistic model to the results of the matches (in terms of the distribution of the total number of goals scored in each match).

# 2. DATA SOURCE

Between June, 11 2021 and July, 11 2021, the final phase of the 60[th] edition of the European Football Championship, called UEFA EURO 2020 even though the matches were played in 2021, took place. A total of 51 matches were played: 36 in the *group stage* and 15 in the

*knockout phase* (round of 16, quarter-finals, semi-finals and final). All the matches in the group stage lasted 90 minutes, while, for the matches in the knockout phase, if a match was level at the end of normal playing time, extra time was played for a total of 120 minutes (90 minutes of normal playing time plus 30 minutes of extra time). The times at which the goals were scored were retrieved from the official UEFA website (https://www.uefa.com/uefaeuro-2020/).

Data were analysed employing R, version 4.1.1 (R Core Team, 2021).

## 3. ANALYSIS OF THE NUMBER OF GOALS (NORMAL PLAYING TIME ONLY)

Considering only the normal playing time of the 51 matches, a total of 135 goals were scored (94 in the group stage and 41 in the knockout phase). The average number of goals was 2.65/match.

There is a good agreement between observed and expected frequencies calculated employing the Poisson distribution with mean $\mu$ equal to the observed one ($\chi^2 = 0.753$; d.f. = 5; $p = 0.98$).

There are other tests, more specific for the Poisson distribution (Rao and Chakravarti, 1956).

It is well known that for the Poisson distribution with parameter $\mu$, the mean as well as the variance are both equal to $\mu$. Therefore a "quick" way to evaluate if a Poisson distribution can fit the data is a comparison between the mean and the variance of the data at hand. In our case these values are, respectively, 2.65 and 2.43.

In particular, Fisher (1950) discussed "the special test for discrepancy of the variance" i.e. the *dispersion index D*. This index is widely used to test for homogeneity of the observations and is also referred to as the variance test for homogeneity.

In our case, the index of dispersion is $D = 45.96$ (with 50 degrees of freedom) with an associated *p-value* of 0.636, which is clearly not significant.

## 4. EXTRA TIME AS A SEPARATE MATCH

During the knockout phase, there were 8 matches that ended normal playing time with a draw. In these cases, extra time was played (two periods of 15 minutes each). The total number of goals scored in the extra times was 7, i.e. a mean of 0.875/extra time. Dividing this value by the extra time period (30 minutes) we obtain a mean of 0.0292 goal/min, quite similar to the corresponding value found when considering only normal playing time (0.0294).

Considering only the 7 goals scored during the extra time, there were 4 matches without goals (France - Switzerland, Switzerland - Spain, Italy - Spain, Italy - England), 2 matches with 1 goal (Sweden - Ukraine, England - Denmark), 1 match with 2 goals (Croatia - Spain) and 1 match with 3 goals (Italy - Austria).

In this case the large sample approximation of the dispersion index $D$ may not be ideal. Therefore, using the algorithm proposed by Frome (1982), we employed Fisher's exact variance test for the Poisson distribution, which is based on the (conditional) exact sample distribution of $D$ (Fisher, 1950). The exact *p-value* associated with the observed dispersion index (10.14), given the total number of goals observed (7), is 0.246, giving no clear evidence for a deviation from a Poisson distribution.

## 5. ANALYSIS OF ALL THE GOALS SCORED IN THE 51 MATCHES

As mentioned above, in the final phase of UEFA EURO 2020 there were 43 matches which ended after normal playing times (36 in the group stage and 7 in the knockout phase), while in the remaining 8 matches extra time was necessary. In this section, we consider the number of goals scored in each match irrespective of its duration.

Fisher's exact variance test for the Poisson distribution to the number of goals scored gave results that were not significant ($p$ = 0.813 for the 43 matches lasting 90 minutes and $p$ = 0.242 for the 8 matches lasting 120 minutes).

The chi square test for the goodness of fit applied to the number of goals scored in each match irrespective of its duration was 1.032 with an associated *p-value* greater than 0.9. Therefore, when also considering the total duration of each match, the Poisson appears to model the number of goals scored quite well.

## 6. ANALYSIS OF THE TIMES BETWEEN GOALS

Another way of evaluating the course of a match is to measure the waiting time between two successive goals, or between the start of the match and the first goal. A distinctive aspect of this type of analysis should be mentioned here: the presence of censored times. In fact, if a game ends without goals scored, it is not possible to measure the time elapsed between the start of the match and the first goal. In this case we define the time measured (90 minutes or 120 minutes in the case of extra time) a censored time. Similarly, if the last goal was scored after e.g. 75 minutes and there was no extra time, we consider a censored time of 15 minutes.

In the case of UEFA EURO 2020, we observed a total of 142 goals during a total of 4830 minutes (considering both normal playing and extra time); therefore, the average rate was 0.029 goal/min. The reciprocal of this value (i.e. 34) has the dimension of min/goal, and can be considered a measure of how many minutes elapse, on average, between one goal and the next.

Considering the 142 waiting times and taking into account also the censored times, we can estimate, the survival curve associated with these times by employing the product limit estimator proposed by Kaplan and Meier (1958).

In our analysis, we have a total of 188 observations (i.e. times): 142 are waiting times, while the remaining 46 are censored times (there were 5 matches with a goal scored just at the end of the match: Hungary-Portugal, Turkey-Wales, Sweden-Poland, Wales-Denmark, Sweden-Ukraine).

The lowest waiting time was 1 min, which was observed during the quarter-final match between Belgium and Italy: Belgium scored a goal at the end of the first play time (45'), one minute after a goal scored by Italy. The highest observed waiting time was 95 min, occurred during the round of 16 match between Italy and Austria: the first goal of the match was scored by Italy in the extra time, 95 minutes after the beginning of the match. The estimated median time was 24 minutes. Therefore, the estimated probability that one had to wait at least 24 minutes to observe another goal is 0.5.

We have seen that the average rate of goal scoring is 0.029 goal/min. If we consider an exponential distribution $f(t) = \lambda e^{-\lambda t}$, the survival probability $P(T > t)$ is given by $S(t) = e^{-\lambda t}$. If we assign the observed average rate (i.e. 0.029) to $\lambda$, we can compare the (expected) survival curve with the empirical one. There is a good agreement between expected and observed results. The median survival time is $-\log(0.5)/\lambda$, i.e. 23.6 minutes in our case, which is in a fairly good agreement with the observed value (24 minutes).

This result was expected, since it is well known that if the number of events observed during a fixed time period follows a Poisson distribution, then the waiting times between events follow an exponential distribution.

## 7. CONCLUSIONS

Football is one of the most popular and followed sports. The outcome of a match can be thought of as depending on a number of factors such as the ability of the teams, the availability of more or less strong and trained players, the tactics defined by the coaches, etc. In this sense, the result of a match cannot be thought of as a game of chance.

However, if we look at the results of the overall sets of the 51 matches of UEFA EURO 2020 and, more precisely, at the distribution of the number of goals/match, we found that a simple model like the Poisson appears to fit the data quite well.

This result is in line with studies by authors who have found the Poisson model appropriate for analysing the results of football matches (Dixon and Coles, 1997; Karlis and Ntzoufras, 2003; Maher, 1982).

We do not claim that the 51 matches of UEFA EURO 2020 are 51 independent homogeneous Poisson processes with a constant probability per unit of time of scoring a goal (which does not vary from match to match), "thus degrading football to a pure game of chance" (Bittner et al., 2009), since football matches are a system of highly co-operative intercorrelated entities. However, the results obtained with a model based on a single parameter are surprisingly good.

The Poisson is perhaps the simplest distribution for modelling count data. When overdispersion is present in the data, the negative binomial (Kilpatrick, 1977) or the Waring (Canal and Micciolo, 1999) distribution can be employed. However, in UEFA EURO 2020 the variance of the number of goals was lower than the mean.

Harris, Yang and Hardin (2012) considered a model based on the generalised Poisson distribution (Consul and Jain, 1973) which can accommodate both overdispersed and underdispersed count data. In this model the dispersion factor depends on a parameter $\delta$ which can assume positive (overdispersion), null (equidispersion) or negative (underdispersion) values. In the case of UEFA EURO 2020 the estimate of this parameter was −0.055, with an associated 95% confidence interval between −0.25 and 0.15 and a *p-value* for testing the hypothesis $\delta = 0$ of 0.299, giving no evidence against the more parsimonious Poisson model.

The good fit of the exponential distribution to model the waiting times between goals is expected from a theoretical point of view, but could be considered somewhat counter intuitive and unrealistic given the memoryless property of the exponential model.

Underlying the exponential model and the resulting Poisson model is the concept of independence between events, a fundamental of probability theory. Independence can be thought of as a useful way to model outcomes, like the result of tossing a coin or of a game of chance, that are not inherently random but can be predicted exactly (at least in principle) once the relevant parameters are specified. However, in various situations in the real word, people often discard independence, suffering from cognitive illusions (Tversky & Kahneman, 1974); one famous case can be found in basketball and is known as "hot hand" (Gilovich, Vallone & Tversky, 1985).

The surprising fit of the Poisson model can be at least partly explained by considering some features of the case study: the teams participating in the final phase came from a selection phase, so their skills were relatively homogeneous; the time span in which the final phase of the competition took place was short, which prevents the skills of the different teams from greatly improving; in a competition like this no distinction between home or away games can be made;

the abilities of the two teams were averaged by considering only the total number of goals scored in each match rather than counting them separately.

Although UEFA EURO 2020 can certainly not be considered a "pure game of chance", this competition can be seen as another appealing example of the pervasive real-world ubiquity of the concept of independence and its ability to encompass complex situations in a single, extremely parsimonious model.